**Diplomacy and science in the context of World War II: Arthur Compton's 1941 trip to Brazil**[1]

Olival Freire Junior [#]

Indianara Silva [*]

[1] This paper was published in *Revista Brasileira de História*, São Paulo, 34 (67), pp. 181-201, 2014. A condensed version of this work was presented to the panel, "Technoscience Exchanges Between Latin America, Europe, and the United States in the 'Short Twentieth Century': Comparative Studies of Knowledge and Practice Exchanges", at the Annual Meeting of the History of Science Society, Cleveland, Ohio, between 3 - 6 November 2011. One of the authors (IS) would like to thank Washington University in Saint Louis and the American Institute of Physics (AIP), in particular Sonya Rooney and Gregory Good, for their support and grant aid, which enabled the consultation of the Arthur Holly Compton Papers, 1905-1971, deposited in the University Archives, Department of Special Collections, Washington University Libraries, Saint Louis, MO; and the Arthur Holly Compton research notebooks, 1919-1941, at AIP. The authors are thankful to Joan Bromberg, John Krige, Henrique Altemani, Magali Sá, Antonio Videira, Ana Andrade, Amilcar Baiardi, and referees and editor of the *Revista Brasileira de História*, Alexandre Fortes, for their comments on a preliminary version of this work; and to the CNPq, CAPES, FAPESB, and UEFS, without whose support this research would not have been possible.

[#] Universidade Federal da Bahia, Programa em Ensino, Filosofia e História das Ciências. Instituto de Física da UFBA. olival.freire@pq.cnpq.br

[*] Universidade Estadual de Feira de Santana. Doutora em Ensino, Filosofia e História das Ciências (UFBA e UEFS). isilva@uefs.br

**Abstract**

Historical literature has traced US efforts to bring Latin American countries over to the side of the Allies in WWII. Following a period of hesitation and negotiation with the US and Germany, the Brazilian government aligned itself with the US. What remains virtually unknown is the involvement of scientists in these diplomatic arrangements. Recently unearthed material has shown that Arthur Compton's 1941 trip to a scientific conference on cosmic rays in Brazil was linked to efforts led by the Office for Inter-American Affairs under Nelson Rockefeller. For the Brazilian physicists, who, under the leadership of Gleb Wataghin, had excelled in this area of research during the late 1930s, science was a strong motivation for the visit; it was not, however, devoid of political connotations and professional interests.

**Introduction**

The intensity of the cultural and scientific hegemony of the United States, in particular in relation to Latin America, in the second half of the 20th century, necessarily leads observers to question how this relationship could have been established. In other words, it is a power relationship, which was created historically and should not be naturalized as if it were an inevitable consequence, for example, of a supposed Pan American identity or geographical determinism. To illustrate this historical construction, we may remind ourselves of the case of the creation of Brazilian universities. In the 1930s, the founders of the University of São Paulo (*Universidade de São Paulo*: USP), the Federal District University and the University of Brazil, sought to attract qualified teachers from Europe, in particular from France, Italy and Germany.[2] They did not think of seeking such resources in the United States. Following the Second World War, however, the United States was transformed into a Mecca for the training of Brazilian researchers, although training in European countries was still, secondarily, considered.

      It appears, then, that the construction of this North American hegemony must be considered as a process resulting from the outcome of the Second World War and the polarization of the Cold War that followed it. In fact, at least in terms of the hegemonic scenario of North American science set against that of the European

---

[2] SCHWARTZMAN, Simon. *Um espaço para a ciência*: A formação da comunidade científica no Brasil. Brasília: Ministério da Ciência e Tecnologia, 2001, 154-179. This book is a Portuguese translation of SCHWARTZMAN, Simon. A space for science: the development of the scientific community in Brazil. University Park: Pennsylvania State University Press, 1991.

Continent, this was the well-articulated thesis of the historian John Krige in his *American Hegemony and the Postwar Reconstruction of Science in Europe*.[3] In this work, however, we will argue that, at least in the Brazilian case, the hegemony which will coalesce after the Second World War was prepared and anticipated by procedures of cultural diplomacy undertook by the United States between the eve and the beginning of the engagement of that country in the Second World War. If cultural hegemony came as outcome of the war, it was early prepared by the American cultural diplomacy.[4] We will also argue that in this process the Brazilian elite, particularly the scientific elite, actively sought out this realignment, demonstrating that this was not a process imposed on Latin America from the outside. Certainly there was a different scenario during the period following the Second World War.

The first part of this thesis is analogous with that defended by Gerson Moura, in his *Tio Sam chega ao Brasil: A penetração cultural Americana*, when he states that "to be more exact, the visible arrival of Uncle Sam in Brazil really occurred at the beginning of the 1940s, under well-defined conditions and with a clear purpose." The second part, however, moves away from that author's position, since he discusses the

---

[3] KRIGE, John. *American Hegemony and the Postwar Reconstruction of Science in Europe*. Cambridge: The MIT Press, 2006.

[4] For the concept of cultural diplomacy, see CULL, Nicholas J. Public Diplomacy: Taxonomies and Histories. *The ANNALS of the American Academy of Political and Social Science*, v. 616, p. 31-54, 2008. American cultural diplomacy in the 20th century is presented in ARNDT, Richard T. *The First Resort of Kings – American Cultural Diplomacy in the Twentieth Century*. Washington, DC: Potomac Books, 2005. In this book, one of the chapters, pp. 75-97, presents the character and the period we are interested in this paper: Nelson Rockefeller, 1940-1945.

North American cultural influence on Brazil that developed during this period in terms of "ideological penetration and market gains", thus, unlike our thesis, attributing a passive role to the Brazilian elite who participated in this process. In the same way, the first part of the thesis coincides with the position of Antonio Tota in his book *O imperialismo sedutor: A americanização do Brasil na época da Segunda Guerra*. The title of Antonio Tota's book (*Seductive Imperialism*) is loaded with the metaphor of seduction, suggesting a certain passivity on the part of the Brazilian elite protagonists. This impression is only dissipated in the book's final chapter, when the author discusses how "our Americanization obviously did not happen passively."[5]

Our argument will be developed by analysing the activities of the Office of the Coordinator of Inter-American Affairs (OCIAA), a body of the North American State run by Nelson Rockefeller, which specifically functioned between 1941 and 1945, related to the visit of a North American scientific committee to Brazil. This committee was led by the physicist Arthur H. Compton, from the University of Chicago, winner of the 1927 Nobel Prize for Physics, with the aim of participating in research and in a conference about cosmic rays held in Rio de Janeiro in 1941. While Compton's visit is well documented in the literature on the history of physics, historians have overlooked his diplomatic connection.

This work poses historians of science with the challenge of overcoming the obstacle of locating science in history rather than merely writing a history of science.

---

[5] MOURA, Gerson. *Tio Sam chega ao Brasil*: A penetração cultural americana. 3ª edição. São Paulo: Editora Brasiliense, 1986, pp. 8 and 11. TOTA, Antonio Pedro. *O imperialismo sedutor*: A americanização do Brasil na época da Segunda Guerra. São Paulo: Companhia das Letras, 2010, p. 191.

To explain more clearly - we are dealing with two relatively well-known processes, which have not previously been connected. On the one hand, diplomatic relationships between the United States and Brazil between 1935 and 1945, which led to Brazil aligning itself with the Allies and to its entry into the Second World War. This process has been well documented, including in classic works, such as in the books of Gerson Moura and Luiz Alberto Moniz Bandeira.[6] Despite the existence of this literature, however, the activities of the OCIAA, whose Brazilian committee was run by the businessman and political advisor Berent Friele, have received scarce, if any, attention from historians focusing on the history of relationships between Brazil and the United States. On the other hand, we have the flowering of research into the physics of cosmic rays at the University of São Paulo, under the leadership of the founder of its Department of Physics, the Russian-Italian Gleb Wataghin. Historians have paid attention to this process, drawn by Brazilian physics' uniquely rapid acquisition of a reputation for international excellence. The generation that was thus trained includes Cesar Lattes, who, even today, is the Brazilian experimental physicist with most extensive international profile. The connection between these two processes, however, has been overlooked both by historians of science and by historians *tout court*, and this is the aspect which our study intends to explore.

---

[6] MOURA, Gerson. *Autonomia na Dependência*: A política externa brasileira de 1935 a 1942. Rio de Janeiro: Editora Nova Fronteira, 1980; BANDEIRA, Luiz Alberto Moniz. *Presença dos Estados Unidos no Brasil*. Rio de Janeiro: Civilização Brasileira, 2007. For a review of the literature, see, ALVES, Vagner Camilo. *O Brasil e a Segunda Guerra Mundial*: História de um envolvimento forçado. Rio de Janeiro: Editora PUC Rio e Editora Loyola, 2002. See also FAUSTO, Boris (ed.). (2007). *História Geral da Civilização Brasileira*: O Brasil republicano. Tomo III, Volume 1. Rio de Janeiro: Bertrand Brasil, 2007; and CERVO, Amado e BUENO, Clodoaldo. *História da política exterior do Brasil*, 3a Ed. ampliada, Brasília: Editora UnB, 2010, pp. 233-267. For a comparison between Brazil and Argentina in the same period, see FAUSTO, Boris e DEVOTO, Fernando. *Brasil e Argentina – um ensaio de história comparada (1850-2002)*. São Paulo: Editora 34, 2004. The proximity between Brazil and the United States, however, dates back to the beginning of the 20th century, an observation for which we would like to thank Henrique Altemani. On this theme, see RICUPERO, Rubens. O Brasil, a América Latina e os EUA desde 1930: 60 anos de uma relação triangular. In ALBUQUERQUE, José Augusto Guilhon (org). *Sessenta Anos de Política Externa Brasileira (1930-1990):* Crescimento, Modernização e Política Externa. São Paulo: NUPRI-USP / Cultura Editores Associados, 1996, p. 37-60. For an examination of this approach, under the aegis of Rio Branco, and its influence on scientific congresses, see SUPPO, Hugo Rogélio. Ciência e relações internacionais – o congresso de 1905. *Revista da SBHC*. Rio de Janeiro: SBHC, 1, p. 6-20, 2003.

Before following with our narrative, let us comment a little further the scant attention in the literature to the activities of the OCIAA and its Brazilian committee. Positive exceptions to this lacuna are the works of Gerson Moura and Antonio Tota. In the booklet *Tio Sam chega ao Brasil*, published in 1986, Gerson Moura took up the theme again, looking at the role of Berent Friele as Coordinator of the OCIAA's Brazilian committee. Although an important work, given its uniqueness in national literature, in our opinion this booklet does not adequately explore the primary sources and treats the themes of science and education in these relationships very generically. The book *O imperialism sedutor*, written by Antonio Tota, however, adequately explores primary sources, in particular the North American archives, and makes an essential historiographical contribution to the theme. Unfortunately, for our purposes, however, Tota pays little attention to the spheres of education, science and technology, dedicating only two pages to this. Lira Neto, in the recently published book, *Getúlio 1930 – 1945*, notes the OCIAA's activities, Rockefeller heading it, and its support to Walt Disney's journey to Brazil as well as its support to the medical training of Vargas' brother, Lutero.[7] One conspicuous gap is Luiz Alberto Moniz Bandeira's book, *Presença dos Estados Unidos no Brasil*, in which no reference is made to the OCIAA; Nelson Rockefeller appears only as a businessman and not a diplomat, while Berent Friele is an illustrious absence. In an earlier book, *Autonomia na Dependência: A política externa brasileira de 1935 a 1942*, published in 1980, Gerson Moura drew from the secondary literature, that is, from Gerald Haines' study, which, more than forty years ago, stated that, "While much has been written on the political and economical efforts of the Roosevelt administration to ensure the safety of the American continents during this period, little attention has been focused on the far more subtle attempts by Roosevelt and his advisers to promote hemispheric cooperation and unity and to make the New World an American hemisphere." As far as we have been able to review the literature, the Haines study was not followed up. Thus, the archives relating to the OCIAA, as relevant as they may be for the history of Latin America, remain a territory as yet under-explored by historians.[8]

---

[7] The positive exceptions are (MOURA, 1980, pp. 139-141), (MOURA, 1986) and (TOTA, 2010). Science and education in the relationship Brazil- USA is approached in (MOURA, 1986, pp. 47-50) and (TOTA, 2000, pp. 80-82). NETO, Lira. Getúlio 1930-1945: Do governo provisório à ditadura do Estado Novo. São Paulo: Companhia das Letras, 2013, pp. 397 e 468.

[8] (BANDEIRA, 2007). (MOURA, 1980). (HAINES, Gerald. Under the Eagle's Wing: The Franklin Roosevelt Administration Forges an American Hemisphere, *Diplomatic History*, Hoboken: Wiley-Blackwell, 1, 4, p. 373-388, 1977, p. 373).

The work is organized in the following way: the first part is dedicated to a brief review of the context and literature regarding Brazilian diplomacy throughout the 1930s and during the Second World War, which constitutes the historical background of Compton's visit in the mid-1941; while the second, which results from original historical research, explores the connections between the scientific and political aspects of Arthur Compton's visit to Brazil in 1941. In this part, information about the political positions of Paulus Aulus Pompéia and Gleb Wataghin throw light on as yet unexplored aspects of the literature regarding the history of science. In the conclusion, we draw certain implications for the historiography of science.

**1. The Brazil - Germany – United States triangle in the 1930s**

Brazil's entry into the Second World War in 1942 on the side of the Allied forces was the result of one of the most complex processes of our diplomatic history. The Vargas government, a product of the 1930 revolution, and from 1937 a dictatorship resulting from the New State's coup d'état, contained disparate tendencies in a world polarized by the ascension of Nazism in Germany in 1933. The government accommodated both modernizing projects and pro-fascist tendencies. Modernizing projects came from the *tenente* movements of the 1920s, resulted in the developmentalist trends in Brazilian politics, and combined with liberal tendencies, personified by Oswaldo Aranha, minister of Foreign Affairs. The pro-fascist tendencies were represented by Generals Eurico Gaspar Dutra and Góis Monteiro. Throughout the 1930s, Vargas sought to exploit the international polarization to advance his own agenda, which specifically involved the construction of a large-scale steel mills and the modernization of armaments for the armed forces.[9] According to Cervo and Bueno, diplomatic historians, "in historiography there is a certain consensus that Brazil played a 'double game' in relation to the United States and Germany, in the period leading up to the Second World War, using it as a bargaining position. This game was facilitated by growing German participation in Brazil's foreign trade in the period between 1934 and 1938, at the same time as a decline in both the North American and English presence in the country's imports and exports."[10]

If this was the Brazilian game, then what game did the other actors, in particular the United States and Germany, play during the 1935 – 1941 period of Brazilian foreign policy, which Moura (1980) described as "pragmatic equidistance"? Nazi Germany, willingly accepting Vargas Government policy and exploiting the legal framework of the Brazil-Germany Compensation Adjustment, became Brazil's largest trade partner, overtaking England and the United States and using economic exchange for political approaches, both to government members and to *integralistas*, from the fascist-inspired Brazilian party, as well as to German immigrants settled in the south of the country. The outbreak of war in 1939, however, limited the scope of German manoeuvres, as a specific result of the naval blockade imposed by England on German-Brazil trade. Thus, only a portion of the armaments bought from Krupp arrived in Brazil and supplies for the construction of the steel mills sought by Vargas were promised for after the war.[11]

---

[9] LEVINE, Robert M. *Father of the poor?* Vargas and his era. Cambridge, MA: Cambridge University Press, 1998; LEVINE, Robert M. & CROCITTI, John J. (eds), *The Brazil Reader*: history, culture, politics. Durham: Duke University Press, 1999; HENTSCHKE, Jens R. (ed.), *Vargas and Brazil*: new perspectives. New York: Palgrave Macmillan, 2006. SKIDMORE, Thomas E. *Politics in Brazil, 1930-1964*: an experiment in democracy. New York, Oxford University Press, 1967.

[10] (CERVO e BUENO, 2010, p. 234)

At the beginning, the Americans were tolerant with trade between Brazil and Germany, even making economic offers to Brazil. According to Moura, "all these concessions indicate that there was no absolute United States hegemony regarding the continent in general and Brazil in particular."[12] The evolution of the international political situation, particularly following the Spanish Civil War in 1936 and the outbreak of war in Europe in 1939, led the USA to prepare for a possible war scenario. In this context, Latin America, and Brazil in particular, enjoyed greater importance in American foreign policy. This is related to the consolidation of its influence in the hemisphere, at the same time neutralizing the influence of Axis countries in the region, in particular of Germany and Italy. Brazil, specifically, acquired strategic military importance, given that the northeast projection around the city of Natal supplied a natural platform for North American aviation to fly to North Africa and Europe. For this reason, while Moura calls the 1935 – 1937 period "possible equilibrium" and the period between 1938 and 1939 "difficult equilibrium," the subsequent period was called "the ideological and political offensive by the United States (1939 – 1942)." For the purposes of this work, we are particularly interested in this ideological offensive. Based on the work of Gerald Haines, Moura described this North American offensive in the following manner:

---

[11] (CERVO e BUENO, 2010, pp. 253-7)

[12] (MOURA, 1980, pp. 98-9),

> But 'Pan American' values did not propagate spontaneously, nor were the continental conferences a sufficient instrument for their dissemination. For this reason, the United States government adopted a series of measures to guarantee the success of the undertaking. This was to give new encouragement to the already existing Pan American institutions and to American government bodies, such as the Cultural Division of the State Department and the Interdepartmental Committee on Cooperation with the American Republics. Perhaps the more important decision was the creation of OCIAA (Office of the Coordinator of Inter-American Affairs) under the direction of Nelson Rockefeller and charged with counter-balancing growing Axis propaganda in Latin America. The extent of this new body's activities may be assessed by examining its aims: to persuade the Latin American nations to follow the leadership of the USA in opposition to the Axis, to integrate the American economy with that of the USA into one system, to prevent revolutions in the Americas, to fight against Axis agents and to increase United States trade. And all this without giving the impression that they were interfering in the internal business of sovereign States.[13]

The pinnacle of the "double game" played by Vargas was the speech he made in June 1940 aboard the ship Minas Gerais, in which he signalled the possibility of Brazil aligning itself to the Axis powers, if this were to be a pathway for heavy industrialization, meaning the construction of the steel mills. According to Cervo and Bueno, "the international situation was favourable. With the advance of the conflict, this was the right moment: either Washington cooperated, straightaway, or a pathway to German cooperation would be opened up." Germany offered to construct the steel mills when the war was over. The speech frightened the North Americans, who speeded up government decisions to offer Vargas what he wanted, but negotiations were not yet concluded. The precipitation of political and military events, in the following period, accelerated these undertakings. In December 1941, Japan attacked Pearl Harbour, dragging the United States into the world war. In January 1942, the Pan American Meeting was held in Rio de Janeiro; here the United States expected a rupture in diplomatic relations on the part of the Latin American states in relation to the Axis powers. There were obstacles, however, since Chile and Argentina defended a position of neutrality in face of the current conflict. During the meeting, Vargas negotiated directly with Roosevelt for a guarantee of the construction of what would become the Volta Redonda steel mills and the rearmament of Brazilian military forces. Vargas obtained both. With this commitment in hand, at the end of the meeting the Brazilian president announced a rupture of diplomatic relations between Brazil and the Axis powers.

---

[13] (HAINES, 1977, pp. 379-380), as cited by (MOURA, 1980).

According to Fausto and Devoto, in a comparative analysis of Brazilian and Argentinian foreign policies, "It is evident that the Vargas option revealed, at the same time, his non-ideological strategy and his political acumen, since alignment with the Allies took place when the Axis was still militarily victorious."[14] The military benefit for the United States was extremely significant. It is estimated that, now with a modernized airport and a North American military presence, more than 25,000 North American planes passed through Natal. The rupture of relations made Brazil a target for German retaliations, with the sinking of several Brazilian ships.[15] The epilogue of this history is well known. Under the pressure of public opinion, which included both liberals and the imprisoned left, Vargas declared war on the Axis powers in August 1942, and this was followed by the preparation of the Brazilian Expeditionary Force, which fought in Italy on the side of Allied forces.[16] This epilogue forms part of our national and international political memory. In 2011, on visiting Brazil, the British Deputy Prime Minister began his statements by remembering that "[d]uring the Second World War, Brazilian soldiers fought alongside our forces in Italy."[17]

---

[14] (CERVO e BUENO, 2010, p. 260); (FAUSTO e DEVOTO, 2004, pp. 272-8). Our narrative approaches that of (MOURA, 1980) and diverges from (ALVES, 2002) to the extent that the latter eliminates the autonomy, however relative, of Vargas' political activities, reducing them to a mere reflection of existing power relations.

## 2. Compton's visit to Brazil in 1941: science and diplomacy

From the outset, approaches between the groups led by Arthur Compton in the United States and Gleb Wataghin in Brazil, contained a strong element of scientific collaboration of interest to both sides. Until the end of the 1920s, Compton had worked on X-ray scattering, work that earned him the Nobel Prize for Physics in 1927. After this, Compton changed direction, dedicating his time throughout the 1930s to the study of cosmic rays. Through this study, he became involved in a heated controversy with Robert Millikan, also a Nobel Prize winner, about the nature of cosmic rays. The controversy between these two physicists could have been resolved with the help of precise measurements of cosmic rays at several latitudes. If X-rays were highly energetic photons, and therefore had no charge, as Millikan advocated, they would not suffer variations on interaction with the Earth's magnetic field. If, on the other hand, they were charged particles, different angles of incidence with the Earth's magnetic field at different latitudes would lead to different interactions. Compton and Millikan actively sought out these measurements. Compton, in particular, organized an expedition at the beginning of the 1930s to undertake measurements at various latitudes; this included an excursion to Latin America. The results revealed the latitude effect and Compton, in a dispute covered by the North American mainstream media, won the argument.[18] Throughout the 1930s, Compton continued to improve his measurements, despite the resolution of this controversy. In the second half of the decade, cosmic rays promised to shed light on another enigma of physics. In 1934, the Japanese physicist Hideki Yukawa, in a work that won him the Nobel Prize, introduced the idea of a new force of nature, the strong or nuclear force, to explain the stability of the nucleus. Yukawa's hypothesis also involved the existence of a new particle, called the mesotron, later called the meson, an intermediate mass between the electron and the proton, which was the mediator of this new force.  Although the 1930s had seen the construction of the first particle accelerators, physicists' hopes remained focused on cosmic rays, which were then more energetic than the particles produced in accelerators. What made this subject even more intriguing was the experimental data at the end of the 1930s, which appeared to suggest the existence of more than one mesotron.  In the 1940s, Compton therefore maintained his interest in something which had been his preferred subject for more than ten years, although already competing with the work he was doing with a number of other physicists at the University of Chicago, including Enrico Fermi, on nuclear fission; work which would later result in the Manhattan Project and the construction of the first atomic bomb.

---

[15] See (SANDER, Roberto. *O Brasil na mira de Hitler*: A história do afundamento de 34 navios brasileiros pelos Nazistas. Rio de Janeiro: Objetiva, 2011), a work notable for the quality of its narrative.

[16] For this support from the left, see Roberto Sisson's letter to Vargas at the beginning of 1942, cited in (MOURA, 1980, 175). Sisson was one of the leaders of the ANL, the left political organization which was proscribed following the failed communist uprising of 1935.

[17] CLEGGE, Nick. *Folha de São Paulo*, 20 June 2011.

---

[18] DE MARIA, M. and RUSSO, A. Cosmic Ray romancing: the discovery of the latitude effect and the Compton – Millikan controversy. *HSPS*, Berkeley: University of California Press, 19(2), 211-266, 1989.

Gleb Wataghin was the founder of the Physics Department at the University of São Paulo.  Russian by birth, he had immigrated to Italy at the end of his adolescence, and there he concluded his secondary education, studying Physics and becoming a teacher at the Polytechnic of Turin.  In 1934, the founders of the University of São Paulo invited him to lead the Physics Department at the recently created university.  A little later on, in 1938, another Italian physicist, Giuseppe Occhialini, who had worked with P.M.S. Blackett in England, entered the same department.  Wataghin knew how to surround himself with talented and promising young people, such as Marcelo Damy de Souza Santos and Paulus Pompéia, the experimentalists, and Mário Schenberg the theorist, and how to choose a research topic – cosmic rays – that was relatively cheap but could, if it worked well, place the young São Paulo physics at the forefront of world physics.[19] The research bore fruit, and at the end of the 1930s, they observed what at the time were called "penetrating showers," that is, the production of various particles following the interaction of cosmic rays at the higher levels of the atmosphere. Such results were significant for the study of mesotrons, or mesons, as these particles came to be called. The São Paulo group took measurements at high altitudes, in aeroplanes and using balloons, and at depths of about 30 metres below the surface of the Earth. Damy developed electronic techniques, which enabled them to gain accuracy in the measurements.[20] As we shall see, it was these results that attracted Compton's notice and shaped his interaction with the São Paulo physicists.

---

[19] (SCHWARTZMAN, 2001). COSTA RIBEIRO, Joaquim. A Física no Brasil. In AZEVEDO, Fernando (ed.). *As ciências no Brasil*, Vol 1. 2ª ed. [1ª ed. 1955]. Rio de Janeiro: Editora UFRJ, 1994, pp. 191-231. MOTOYAMA, Shozo. História da Física no Brasil. In MOTOYAMA, Shozo e FERRI, Mário Guimarães (Eds.). *História das ciências no Brasil*. São Paulo: EPU e EDUSP, 1979. BUSTAMANTE, Martha CECILIA. Giuseppe Occhialini and the history of cosmic-ray physics in the 1930s: From Florence to Cambridge and ANDRADE, Ana Maria Ribeiro. Occhialini's trajectory in Latin America. In REDONDI, Pietro et al (Eds), *The scientific legacy of Beppo Occhialini*. Bologna: Italian Physical Society, 2006, pp. 35-49 and 51-69, respectively. VIDEIRA, Antonio Augusto Passos & BUSTAMANTE, Martha Cecilia. Gleb Wataghin en la universidad de São Paulo: un momento culminante de la ciência brasileña, *Quipu*, Cidade do México: SLHCT, 10, 3, 1993, p. 263-284.

[20] WATAGHIN, Gleb; SOUZA SANTOS, Marcelo Damy and POMPÉIA, Paulus A. Simultaneous penetrating particles in the cosmic radiation, *Physical Review*, 57, 61, 1940; *idem*, Simultaneous penetrating particles in the cosmic radiation, II, *Physical Review*, 57, 339, 1940. DAMY, Marcelo de S. S.; POMPÉIA, Paulus A. and WATAGHIN, Gleb. Showers of penetrating particles, *Physical Review* 59, 902-3, 1941.

One should also note that Wataghin encouraged both the publication of results in international journals and the international circulation of young Brazilian physicists. Schenberg visited Europe before the outbreak of war, and later, with a grant from the Guggenheim Foundation, went to the United States, where he worked and published on astrophysics with George Gamow and Subrahmanyan Chandrasekhar.[21] Marcelo Damy de Souza Santos took up an internship in Cambridge, England and Paulus Pompéia went to the United States in 1940 to work with Compton, through an internship in Chicago.

With the work about penetrating showers, Wataghin sought to establish scientific collaboration with Compton, which was well received by the North American. Wataghin submitted his results to the North American journal *Physical Review*, but also sent them directly to Compton prior to their publication. "Thank you for sending on to me the manuscript of your 'Letter' to the Editor of the PHYSICAL REVIEW on 'Showers of Penetrating Particles'," acknowledged Compton in 1941. Compton continued to demonstrate evidence of his interest in intra-nuclear particles, as we can see in these experiments: "we also are getting evidence of mesotrons associated with extensive showers."[22] In this exchange of correspondence and through

---

[21] Regarding Schenberg, see the introduction in HAMBURGER, Amélia Império (ed.). *Obra Científica de Mario Schönberg*: Volume 1 - de 1936 a 1948, São Paulo: EDUSP, 2010. On Schenberg and Chandrasekhar, see Davide Cenadelli, "Solving the Giant Stars Problem: Theories of Stellar Evolution from the 1930s to the 1950s," *Arch. Hist. Exact Sci.* 64, 203–267, 2010.

[22] Compton to Wataghin, 4 January 1941. Series 02 Box 017. Folder: South America Expedition, 1941. Arthur Holly Compton Personal Papers. University Archives, Department of Special Collections, Washington University Libraries. (Hereafter

the correspondence between Wataghin and Pompéia, in December 1940, Wataghin was informed that Compton intended to organize a new expedition to South America (something he had undertaken at the beginning of the 1930s) to measure cosmic rays, and was including Brazil, and more specifically São Paulo, as one of the locations for these measurements.[23] Wataghin reacted very positively to news of Compton's expedition by stating that"[i]t was a great pleasure for me and for my co-workers to learn from yours and Mr. Pompeia letter, that you shall probably come to Brazil next summer."[24] Wataghin placed the resources of USP's Physics Department at

---

abbreviated as CP). In another letter, Wataghin declares, "I am sending to you herewith a copy o four letter to the Editor of Phys. Rev. concerning 'Showers of penetrating particles,'" and reveals that he has identified mesotrons in underground measurements of cosmic rays, as well as at high altitudes. Wataghin to Compton, 26 December 1940. Series 02 Box 017. Folder: South America Expedition, 1941. CP. And, further, "I want to thank you also for sending a letter to the Editor of Physical Review, and for the very interesting informations on recent cosmic ray results obtained in your laboratory." Wataghin to Compton, 12 November 1940. Series 02 Box 017. Folder: South America Expedition, 1941. CP.

[23] Compton to Wataghin, 4 January 1941, Series 02 Box 017. Folder: South America Expedition, 1941. CP. In this letter, it was evident that the choice of São Paulo for the measurements which Compton was planning on South America's Atlantic coast came from a suggestion made by Pompéia: "For this [balloon experiments near the magnetic equator] on advice of Pompeia we have tentatively selected the State of Sao Paulo in southern Brazil as a location where we can hope to recover the balloons which are sent up."

[24] Wataghin to Compton 31 December 1940, Series 02 Box 017. Folder: South America Expedition, 1941. CP. And, further, "I was very glad to learn from Pompeia's letter that you and Prof. Hilberry, Jesse, Hughes and Wollan are coming to Brazil" Wataghin to Compton, 12 April 1941, Series 02 Box 06. Folder 3, CP. This

Compton's disposal and mobilized the university administration and the São Paulo government for logistical support, particularly for the measurements undertaken in balloons, which were carried out in the state of São Paulo. As well as these initiatives, Wataghin sought support from the federal government in Rio de Janeiro and, with the support of the President of the Brazilian Academy of Sciences, Arthur Moses, began to organize a Symposium about Cosmic Rays, to be held in Rio de Janeiro at the end of Compton's visit. This symposium was clearly a homage to Compton's arrival, since the most important scientific discussions would take place in São Paulo before Compton went to Rio.[25] As a follow up to these activities, Compton received an official invitation from Oswaldo Aranha, the Minister for External Affairs, for his visit and that of his committee.[26] Compton's visit was surrounded by success, as seen in the later publication of the symposium annals; today this symposium is a well-documented event in Brazil's history of physics.[27]

    If Compton's expedition to Brazil was highly successful in scientific terms, its political and diplomatic success appears to have been even greater. As soon as news

---

letter ends with a debatable joke: "they will enjoy the trip, the Brazilian forests and the snakes (in captivity)."

[25] "Apparently the serious scientific conference will be here in São Paulo beginning the 30th since Wataghin says there are few physicists in Rio [...] the time there will be spent mostly in official entertaining." W.P. Jesse to Compton, 20 July 1941, Series 02 Box 017. Folder: South America Expedition, 1941. CP.

[26] Oswaldo Aranha to Compton, telegram, 28 April 1941; and Compton to Oswaldo Aranha, letter, 7 June 1941, Series 02 Box 06. Folder 3. CP

[27] *Symposium sobre raios cósmicos*. Rio de Janeiro: Imprensa Nacional, 1943. (BUSTAMANTE, 2006). (ANDRADE, 2006).

of the planned expedition began to circulate in the United States, Stephen Duggan, Director of the Institute of International Education, realized the political implications it could have, particularly as part of North American efforts to win over Latin American hearts and minds to the Allied side.  Thus it was that on 29th April 1941, he told Compton about an initiative to seek support for the expedition from Nelson Rockefeller: "I immediately wrote to Dr. Robert Caldwell, the associate of Nelson Rockefeller in charge of cultural relations with Latin America. I told him of our fine conversation and expressed the hope that he would get in contact with you to send you as a representative of our country under the auspices of the Coordinator's Committee."[28]  Compton also quickly understood these implications and reformulated the expedition in order to be able to take more physicists and their wives.  This led him to request an increase in the endowment promised by the North American government, from 5,000 dollars to 7,500, despite having declared that each individual would cover their wife's expenses, and later added 3,000 dollars, so that the University of Chicago did not pay all the equipment expenses.[29]  The presence of the wives was part of the diplomatic mission, as confirmed by Compton's assertion to R.G. Caldwell, from MIT and Rockefeller's advisor, "in view of the cultural relations objectives of the expedition, I have encouraged the married members of the party to take their wives with them, [...] From our experience in Mexico and from our contacts

---

[28] Stephen Duggan to Compton, 28 April 1941, Series 02 Box 017. Folder: South America Expedition, 1941. CP.

[29] Compton to R. G. Caldwell, 10 February 1941; R. G. Caldwell to Compton, 28 April 1941, Series 02 Box 017. Folder: South America Expedition, 1941. CP

with Latin-Americans in this country, we believe that this will be of real advantage."[30] Furthermore, Compton, because of his warm nature, was considered the right person for such an initiative. As E. Crossett highlighted, "I know of no one who could do a better job on cultural relations than you and your wife."[31] In later documents, Compton suggests that the participation of the OCIAA was even more significant than merely supporting a project already underway, by asserting that "in the summer of 1941, the University of Chicago was requested by the Co-ordinator [sic] of Inter-American Affairs to send a group of expeditions to South America and Mexico to collaborate with the cosmic ray investigators in those countries."[32] Compton fully accepted the double meaning, both scientific and political, of his expedition, by causing news of it to circulate around the USA. According to a letter from W.P. Jesse to a colleague, "[t]he State Department is very keen for us to go as a sort of good will tour so the expedition is as much diplomatic as scientific, although the scientific results are most important just at this time."[33] The expedition's political success may be assessed through Compton's own opinion. In 1942, as part of the proceedings for the reimbursement of expenses to OCIAA, Compton submitted a report in which he confirmed that "the most successful aspect of our South American work with regard to its influence on international relationships seems to have been the Symposium on Cosmic Rays, held under the auspices of the Brazilian Academy of Science at Rio de Janeiro during the first week of August, 1941."[34] The most significant opinion about this political and diplomatic success, however, came from Nelson Rockefeller, the director of the OCIAA, himself. In May 1942, Compton reported to William P. Jesse, who was also part of the expedition to Brazil, "you may be interested that I had a discussion a few days ago with Mr. Nelson Rockefeller who told me that this expedition of ours, according to his reports, was the most successful one that had been undertaken thus far with regard to its stimulation of good relations between the Americas."[35]

---

[30] Compton to R. G. Caldwell, 26 April 1941, Series 02 Box 017. Folder: South America Expedition, 1941. CP.

[31] E. Crossett to Compton, 26 August 1941, Series 02 Box 07. Folder: General (I). CP.

[32] Compton to F. Watson, 21 September 1942, Series 02 Box 07. Folder: General (II). CP.

[33] W. P. Jesse to G. R. Tatum, 23 April 1941, Series 02 Box 07. Folder: W. P. Jesse. CP.

**Figura 1**

Legenda: Compton in Rio de Janeiro. In the first line, from left to right, Wataghin and Compton are the 2$^{nd}$ and the 4$^{th}$. Source: Academia Brasileira de Ciências

If this was the political perspective from the North American side, how did the São Paulo physicists assess the expedition's success? The first issue to consider is that the physicists, particularly Pompéia and Wataghin, were not naive in political terms. The division between the Allies and the Axis powers divided the world, as noted by historian Eric Hobsbawm, who considered the Second World War to be a kind of civil

---

[34] A. H. Compton, "Report to the Coordinator of Inter-American Affairs – University of Chicago South American Cosmic Ray Expedition, 1941, Series 03 Box 02. Folder: South America, 1941-1942. CP.

[35] Compton to W. P. Jesse, 11 May 1942, Series 03 Box 02. Folder: South America, 1941-1942. CP.

war against fascism.[36] Anti-fascism was part of the culture of the students who studied physics in São Paulo in the second half of the 1930s. Amongst the first Italian lecturers invited to set up the University of São Paulo, some were convinced fascists, for example the mathematics lecturer Luigi Fantapié. Legend has it that Mário Schenberg stopped studying mathematics, moving to physics, because of his opposition to Fantapié's fascism.[37] Further, the state of São Paulo was opposed to the Vargas government and its intellectuals were predominantly liberal. Hobsbawm's reference to civil war is significant because it signals that even for many Italians and Germans the struggle against fascism spoke louder than national loyalties. Thus it was that Wataghin did not hesitate to declare to Compton his unequivocal alignment against the Axis powers, which included Italy, where he had been trained, and the Italian government, of which he was an employee, since he maintained professional ties with the University of Turin. In 1942, he wrote, "I cannot say how much I regret not to have the possibility to fight against nazy on the russian front [sic]. I have broken any relation with Italy and have declared to be ready to fight for the United Nations. If my work could be of any use in the U.S.A. I should be happy to go there

---

[36] HOBSBAWM, Eric. *The age of extremes:* The short twentieth century 1914-1991. London: Penguin, 1994. However, Hobsbawm excluded regions under colonial rule, such as Africa, parts of Asia and the Far East, from what he called the "international ideological civil war". This theme is developed further in Chapter 11 of HOBSBAWM, Eric. *How to Change the World*: Tales of Marx and Marxism. London: Little, Brown, 2011, although in this chapter Hobsbawm is more interested in the European and Marxist component of anti-fascism.

[37] Paulus Aulus Pompéia, interviewed by Simon Schwartzman and R. Guedes, CPDOC/FGV, Rio de Janeiro, 1977.

and make some defence work." We remark that Wataghin's anti-fascist position has not been noticed by historians who have written about him.[38] On 19th December 1941, immediately after the United States entered the war, W.P. Jesse finished a letter to Wataghin with a reference to the Soviet Union's resistance to the Nazi invasion – "we are greatly thrilled by the latest Russian news!" –, which highlights the political feelings that the North American and USP physicists shared at that time.[39] Pompéia also clearly identified with Brazilian liberals opposed to the Axis and did not disguise this. Thus it was that, in a letter to the Vice-President of the USA on Pompéia's return to Brazil, Compton presented him as someone on whom the United States could depend for the war effort, not only because of his technical abilities but also his politics: "he has shown his ability persistently to do hard work, skill in administration with attention to details, good understanding of technical problems, a thorough understanding of the difficulties of securing fast action from his own people and a *complete loyalty to the cause of the United Nations.*"[40] Further, judging by Compton's words, he was the principal driver of the expedition to Brazil. On seeking priority transportation from the government of the United States for Pompéia's return,

---

[38] Wataghin to Compton, 8 August 1942, Series 03 Box 02. Folder: South America, 1941-1942. On Wataghin, see (VIDEIRA e BUSTAMANTE, 1993) and PREDAZZI, Enrico. Gleb Wataghin. In ROERO, Clara Silvia (org.). *La Facoltà di Scienze Matematiche Fisiche Naturali di Torino 1848-1998*, 2 Vols., Vol. II, Torino: Deputazione subalpina di storia pátria, 1999, pp. 283-294.

[39] W. P. Jesse to Wataghin, 19 December 1941, Series 02 Box 07. Folder: W. P. Jesse. CP.

[40] Compton to H. A. Wallace, 22 July 1942, Series 02 Box 07. Folder: Cosmic Ray-Meteorology, 1941-1942. CP, our italics.

Compton asserted, "it happens that Dr. Pompeia was the moving spirit which brought about the invitation for our Cosmic Ray mission in the summer of 1941 under the auspices of the Coordinator of Inter-American Affairs."[41]

There is a second aspect to be addressed in order to understand the active attitude of the USP physicists in their support for Compton's expedition. This concerns the most significant scientific collaboration, under Compton's leadership, in the field of cosmic rays. Through this collaboration, the USP physicists expected and obtained support for the supply of equipment for their research, as may be inferred from W.P. Jesse's letter to Wataghin, confirming, "I am enclosing a copy of a letter from the Indiana Steel Company, which they sent me in reply to my inquiries about the magnet that you are interested in. We are also sending a copy of this to the Rockefeller Foundation, which sent us, sometime ago, a list of the things that you need. [...] We are sending to Professor Occhialini some mercury-vapor lamps, and we have on order for him some tungsten lamps."[42] The success of this second aspect was, however, greater than the expectations of the USP physicists, since the reference to the Rockefeller Foundation in the letter cited above deserves expansion. In fact, it could be said that Compton became the principal guarantor for the continued funding that the Rockefeller Foundation secured for the USP Physics Department from this date onwards, and for many years, even after the end of the Second World War. On 23rd January 1942, the Director of the Rockefeller Foundation, H. M. Miller, who had accompanied Compton on his visit to Brazil, wrote requesting his opinion about an application for support made by Wataghin for an expedition to Peru, which had been discouraged by Miller. However, Miller confirms, "although at the time I discouraged a formal request for this purpose, we might be able to reopen the question if assured of the importance of the work to be done ... before taking any step in this direction, we wish to have your considered opinion."[43] As Pompéia remembered much later: "When the war finished we had a very good relationship with Dr Harry Müller Jr [sic] of the Rockefeller Foundation. I knew him in '41 when I came with the American expedition."[44]

---

[41] Compton to J. O. Bell, 14 August 1942, Series 02 Box 07. Folder: General (II). CP.

[42] W. P. Jesse to Wataghin, 19 December 1941, Series 02 Box 07. Folder: W. P. Jesse. CP. In 1945, Wataghin gives thanks, "I received also a letter from Dr. Victor Regener concerning the Geiger-counters. Many thanks for more this help you are giving to our work." Wataghin to Compton, 12 June 1945, Series 03 Box 05. Folder: W, Z. CP.

The support offered by the Rockefeller Foundation to the São Paulo physicists is a theme deserving of more study, which we will develop in another research project. However, we would like to notice how badly the theme has been treated in the literature of history of science in Brazil. Simon Schwartzman, sociologist, In his *Um espaço para a ciência*: *A formação da comunidade científica no Brasil*, describes this support as having begun after the war, although, as we have seen, it began during the WWII and before the US' engagement in the war; Joaquim Costa Ribeiro, in one of the first studies on the history of physics in Brazil, published in the early 1950s, dedicates a very generic paragraph to this theme in terms of dates and institutions;

---

[43] H. M. Miller to Compton, 23 January 1942. Record Group RF RG 1.1 - Projects Series 305 D Brazil. Box 13. Folder 117: University of São Paulo – Physics 1942-1943, Rockefeller Archive Center, Sleepy Hollow, New York.

[44] Paulus Aulus Pompéia, interviewed by Simon Schwartzman and R. Guedes, CPDOC/FGV, Rio de Janeiro, 1977.

Shozo Motoyama, historian of science, in a paper on the history of physics in Brazil, does not address the subject, and Marcelo Damy de Souza Santos, physicist and protagonist of the events we are describing, in his biographical recollections, correctly points out the support from the Rockefeller Foundation to Brazilian physicists from 1940 onwards; however, he incorrectly attributes support for Compton's visit to this foundation, clearly not knowing about the role played by the OCIAA.[45]

**Conclusion**

As asserted at the beginning of this work, the idea of ideological penetration as a result of the OCIAA's cultural initiatives suggests a passive attitude on the part of the Brazilians. References to the creation of the Zé Carioca character by Walt Disney on his arrival in Brazil, as well as the welcome reception given by the Brazilians to one of Hollywood's greatest filmmakers, Orson Welles, when he filmed a documentary concerning Brazilians, both sponsored by the OCIAA, appear to reinforce this passivity.[46] In the cases of the physicists Wataghin and Pompéia, as we have seen, passivity would certainly not be the best way to describe how they acted to forward their professional and political agendas. A more extensive analysis, however, of the way the Brazilians reacted to the OCIAA offensive has yet to be conducted and is beyond the scope of this work. We note, however, that the positive reaction to the North American presence, as stated by the Director of the OCIAA's Brazilian committee, Berent Friele, that, "except among a few disgruntled persons activated by purely personal motives, there is no anti-American feeling left in Brazil," was not maintained in the post-war period, as witnessed by the battles involved in the foundation of Petrobrás, the Brazilian state company for oil exploitation, and the Letter of Testament left by Vargas when he committed suicide in 1954.[47]

---

[45] SCHWARTZMAN (2001, p. 208); (COSTA RIBEIRO, 1994); (MOTOYAMA, 1979) and Marcelo Damy de Souza Santos in *Cientistas do Brasil*, São Paulo: SBPC, 1998, 517-530.

[46] (MOURA, 1980, pp. 139-141). (TOTA, 200). The documentary filmed by Welles was not concluded in his lifetime. Much later, in the 1990s, the film was finally edited, producing the masterpiece "*It's all true*" about the saga of the *jangadeiros* (raft fishermen) who set sail from Ceará to take their grievances to President Vargas.

Finally, seen retrospectively, it does not seem surprising that the OCIAA supported the expedition of a North American scientific leader to Latin America as part of their policy of cultural proximity and their efforts to neutralize the influence of the Axis powers in this part of the Americas. What is surprising is that the theme is only now being addressed in a history article. This brings us to the limitations of the history of science, correctly pointed out by the historian of science John Krige in the opening of his work about North American hegemony in the reconstruction of European science, when he signalled the existence of a "gulf between what diplomatic and economic historians take for granted about the capacity and behavior of the United States to build a world order aligned with its interests and our approach to such an issue (when it occurs to historians of science at all)."[48]

---

[47] Berent Friele, "Report of the Brazilian Division", 25 May 1942, Rockefeller Family, Record Group #4, series Washington DC files – CIAA – Confidential Projects, folder #32, RAC. There is an abundance of literature regarding the episodes after the war, we refer here to (BANDEIRA, 2007, pp. 429-511) and (FAUSTO, 2007).

[48] (KRIGE, 2006, p. 1)